\documentclass[longauth]{aa} 

\usepackage{graphicx}
\usepackage{txfonts}

\usepackage{lineno} 

\usepackage{hyperref}

\usepackage{color}
\definecolor{intreview}{RGB}{0, 0, 0}
\definecolor{extreview}{RGB}{0, 0, 0}

\newcommand{\gamcyg}{PSR J2021+4026}
\newcommand{\gam}{$\gamma$}
\newcommand{\fzero}{$f_0$}
\newcommand{\fone}{$f_1$}
\newcommand{\fermi}{\emph{Fermi}}
\newcommand{\xmm}{\emph{XMM--Newton}}
\newcommand{\degree}{$^{\circ}$}

\newcommand\ireviewme[1]{#1}
\newcommand\ireviewmetwo[1]{#1}
\newcommand\ireviewmethree[1]{{#1}}
\newcommand\ireviewmefour[1]{\textcolor{intreview}{#1}}
\newcommand\jreviewme[1]{#1}

\usepackage{amstext}

\begin{document} 



\title{Multiwavelength observations of \gamcyg\ across a mode change reveal a phase shift in its X-ray emission}


\author{
M.~Razzano$^{1*}$ \and 
A.~Fiori$^{1*}$ \and 
P.~M.~Saz~Parkinson$^{2*}$ \and 
R.~P.~Mignani$^{3,4*}$ \and 
A.~De~Luca$^{3,5}$ \and 
A.~K.~Harding$^{6}$ \and 
M.~Kerr$^{7}$ \and 
M.~Marelli$^{3}$ \and 
V.~Testa$^{8}$
}

\institute{
\inst{1}~Universit\`a di Pisa and Istituto Nazionale di Fisica Nucleare, Sezione di Pisa I-56127 Pisa, Italy\\ 
\inst{2}~Santa Cruz Institute for Particle Physics, Department of Physics and Department of Astronomy and Astrophysics, University of California at Santa Cruz, Santa Cruz, CA 95064, USA\\ 
\inst{3}~INAF-Istituto di Astrofisica Spaziale e Fisica Cosmica Milano, via E. Bassini 15, I-20133 Milano, Italy\\ 
\inst{4}~Kepler Institute of Astronomy, University of Zielona G\'{o}ra, Lubuska 2, 65-265, Zielona G\'{o}ra, Poland\\ 
\inst{5}~Istituto Universitario di Studi Superiori (IUSS), I-27100 Pavia, Italy\\ 
\inst{6}~Los Alamos National Laboratory, Los Alamos, NM 87545, USA\\ 
\inst{7}~Space Science Division, Naval Research Laboratory, Washington, DC 20375-5352, USA\\ 
\inst{8}~INAF Osservatorio Astronomico di Roma, Via Frascati 33, 00078 Monte Porzio Catone, Roma, Italy\\ 
\email{massimiliano.razzano@unipi.it, alessio.fiori@pi.infn.it} \\
}





   \date{Submitted January 9, 2023; accepted May 31, 2023}
   \titlerunning{Multiwavelength observations of \gamcyg}

 
  \abstract
   {We have investigated the multiwavelength emission of \gamcyg, the only isolated \gam-ray pulsar known to be variable, which in October 2011 underwent a simultaneous change in \gam-ray flux and spin-down rate, followed by a second mode change in February 2018. Multiwavelength monitoring is crucial to understand the physics behind these events and how they may have affected the structure of the magnetosphere.
}
   {The monitoring of pulse profile alignment is a powerful diagnostic tool for constraining magnetospheric reconfiguration. 
   \ireviewmethree{We aim to investigate timing or flux changes related to the variability of \gamcyg\ via multiwavelength observations, including \gam-ray observations from \fermi-LAT, X-ray observations from \xmm, and a deep optical observation with the Gran Telescopio Canarias.}}
   {We performed a detailed comparison of the timing features of the pulsar in \gam\, and X-rays and searched for any change in phase lag between the phaseogram peaks in these two energy bands. Although previous observations did not detect a counterpart in visible light, we also searched for optical emission that \ireviewmethree{might have increased due to the mode change}, making this pulsar detectable in the optical.}
   {We have found a change in the \gam-to X-ray pulse profile alignment by 0.21$\pm$0.02 in phase, which indicates that the first mode change affected different regions of the pulsar magnetosphere. No optical counterpart was detected down to $g' = 26.1$ and $r' = 25.3$.}
   {
   \ireviewme{We suggest that the observed phase shift could be related to a reconfiguration of the connection between the quadrupole magnetic field near the stellar surface and the dipole field that dominates at larger distances.} This is consistent with the picture of X-ray emission coming from the heated polar cap and with the simultaneous flux and frequency derivative change observed during the mode changes. 
   }

   \keywords{Gamma rays: stars -- pulsars: individual -- pulsars: variability -- magnetospheres}

   \maketitle
%

\section{Introduction}

The \fermi\ Large Area Telescope (LAT; Atwood et al.\ 2009) sparked a revolution in pulsar astrophysics, with the detection of 
\ireviewmethree{$\sim$300 $\gamma$-ray pulsars}\footnote{\url{https://confluence.slac.stanford.edu/display/GLAMCOG/Public+List+of+LAT-Detected+Gamma-Ray+Pulsars}}.
Among the \fermi-LAT radio-quiet pulsars, \gamcyg\ is particularly interesting for its intrinsically variable behaviour. It was discovered in blind periodicity searches of the \fermi-LAT gamma-ray source coincident with the unidentified EGRET source 3EG J2020+4017 \citep{2009blindsearch}. Deep $Chandra$ observations helped to pinpoint the X-ray counterpart \citep{2011weisskopf}, whose pulsed emission was later detected by \xmm\ \citep{2013lin}. The X-ray phaseogram shows a single broad pulse, while the \gam-ray one shows two peaks separated by $\sim$0.5 in phase.\\ 
\gamcyg\ is seen within the radio shell of the Gamma Cygni (G\, 78.2+2.1) supernova remnant (SNR) in one of the richest and most complex regions of the \gam-ray sky \citep{1977higgs}. 
\ireviewme{Its spin frequency (\fzero\ $\sim$ 3.8 Hz) and frequency derivative (\fone\ $\sim$ $-$8$\times$10$^{-13}$ Hz s$^{-1}$)}
point to an energetic, rotation-powered pulsar (characteristic age $\tau_{c}$ = 77 kyr, spin-down power $\dot{E}_{SD}\sim$ 10$^{35}$ erg s$^{-1}$). Having never been detected in radio \citep{2011ray}, there is no estimate of the pulsar distance from the dispersion measure \citep{2002cordes}, while kinematic models imply a distance of 1.5$\pm$0.4 kpc for the SNR \citep{1980landecker}, yielding an indirect distance estimate for the pulsar based on the assumed association with the SNR. 
The pulsar drew attention when it exhibited a simultaneous change in \gam-ray flux and frequency derivative \fone\ detected by \fermi-LAT in October 2011 (\ireviewme{MJD 55850}) \citep{2013allafort}. \ireviewmetwo{Changes in pulsar frequency derivative have been observed for instance in PSR B0540-69 \citep{2015marshall}, which did not show any simultaneous change in \gam-ray flux, however. \ireviewmethree{This mode changing is still unique}} among the population of \gam-ray pulsars, and therefore, \gamcyg\ has been closely studied. According to \citet{2016Ng}, this event occurred as a consequence of a glitch caused by a rearrangement of the surface magnetic field due to crustal tectonic activity triggered by a starquake that changed the magnetic inclination by $\sim$3$^{\circ}$. \ireviewmethree{The pulsar remained in this state for a few years, then it underwent a gradual recovery phase centred around December 2014} (\ireviewme{MJD 57000}) \citep{2017zhao}, and by August 2015 (\ireviewme{MJD 57250}), it had reached its original flux level and \fone. Following \ireviewme{this recovery}, we triggered a target of opportunity (ToO) observation with \xmm  ~\jreviewme{(PI: Razzano, Obs ID: 118 0763850101)}.
Based on these data, \citet{2018wang} found no significant change in either the pulsar \ireviewme{X-ray} emission or in the relative phase shift between the $\gamma$-ray and X-ray pulse profiles taken around \ireviewme{April 2012 (MJD 56027} (i.e. after the flux drop) and \ireviewme{November 2015 (MJD 57376 (i.e. after the recovery phase}). The pulsar underwent a second flux drop in February 2018 (\ireviewme{MJD 58150}) simultaneously with an increase in \fone\ \citep{2020takata}.\\
Monitoring the relative alignment of the pulse profiles at different wavelengths is a powerful tool for constraining changes in the magnetospheric configuration \citep{2016Ng}; thus, we performed a re-analysis of the two epochs of \fermi\ and \xmm\ data sets to compare the relative phase between the \gam\ and X rays.  Detecting mode changes at other wavelengths can give more insight into the phenomenon, so we also performed and present deep optical observations of \gamcyg\ obtained on \ireviewme{8 June 2016 (MJD 57547)}, that is, shortly after the second \xmm\ observation.\\
\jreviewme{Very few optical observations of \gamcyg\ have been performed to date, and no plausible counterpart has been found.
\ireviewmethree{The first deep observations with the 3.6 m Wisconsin, Yale, Indiana, \& NOAO (WIYN) telescope at the Kitt Peak National Observatory in 2008 reached a detection limit of  $r'\sim 25.2$ and $i'\sim 23$ \citep{2011weisskopf}. New observations performed in 2009 with the 2.2m Isaac Newton Telescope  (INT; La Palma Observatory, Canary Islands) could not detect the pulsar either,  with brighter limits of $B\sim 23.8$, $V\sim 24$ and $R\sim 24$\footnote{The sensitivity of the INT observations, taken with the Wide Field Camera ($32\farcm4 \times 32\farcm4$), was admittedly affected by the ghosts of the very bright star ($V=2.23$) * gam Cyg $\sim 15\arcmin$ arcmin south-east of the pulsar.} \citep{2011collins}. In this paper, we report on} 
} observations that were made in an attempt to identify its optical counterpart and measure its optical flux for the first time.\\
Our paper is structured as follows. In Section 2 we introduce the multiwavelength observations. In Section 3 we describe a new timing analysis of the X and \gam-ray data using two alternative methods, which revealed a change in the relative phase shift between the X and \gam-ray pulse profiles. This result, not found in previous works, implies that the mode change also affects the X-ray emission, suggesting a modification in the magnetospheric regions connecting the dipole field with the inner components close to stellar surface. We have also \ireviewme{reproduced} the results of the previous X-ray timing analyses of the two \xmm\ data sets by \citet{2013lin} and by \citet{2018wang}, respectively. Given the implications of this result, we  discuss all the validations that we performed to rule out possible systematic errors in our analysis. In Section 4 we present the results of our optical observations. Finally, we highlight the theoretical implications of our X and \gam-ray timing results in Section 5 and discuss our plans for multiwavelength monitoring campaigns of this pulsar\footnote{After the February 2018 \gam-ray flux drop, \gamcyg\ underwent a new recovery around June 2020 \ireviewme{(MJD 59010)} (paper in preparation).}. 
Our conclusions are described in Section 6.


\section{Observations}

\subsection{\gam\ -rays}

We analysed $\sim$12 years of \fermi-LAT P8R3 data \citep{atwood2013,bruel2018} from 2008 August 5 (MJD 54683) to 2020 May 26 (MJD 58995). We selected SOURCE class photons of all event types, \ireviewmethree{with zenith angles $\theta_z$<90\degree}. \ireviewme{Photons were taken within 10\degree of \gamcyg\ and with energies E>100 MeV.}
We performed barycentric corrections on the photons using the tool {\tt gtbary} from the Fermi Science Tools v2.0.8\footnote{\url{https://fermi.gsfc.nasa.gov/ssc/data/analysis/documentation/}} and adopted the position of the {\em Chandra} counterpart to \gamcyg\ \citep{2011weisskopf}, $\alpha_{J2000} =20^{\rm h}  21^{\rm m} 30\fs733$; $\delta_{J2000}  = +40^\circ 26\arcmin 46\farcs04$.
\subsection{X-rays}
In this analysis, we considered X-ray photons collected during two \xmm\
observations. 
The first observation (Obs ID: 067059010, PI: Hui) was conducted with a 133 ks exposure on 2012 April 11 (MJD 56028), $\text{about }$6 months after the first mode change, 
when the flux above 100 MeV, $F_{100}$, was (1.10 $\pm$ 0.05) $\times$10$^{-6}$ ph cm$^{-2}$ s$^{-1}$ and the frequency spin-down, \fone\, , was (-8.3 $\pm$ 0.2)$\times$10$^{-13}$ Hz s$^{-1}$ \citep{2018wang}. The MOS1/2 instruments were operated in full-window mode, while the pn was in small-window mode to enable a pulsation search. The second observation (Obs ID: 0763850101, PI: Razzano) was taken with a similar exposure on 2015 December 20 (MJD 57376), $\text{about }$3.7 years after the first observation, when the flux and spin-down of the pulsar had recovered to the initial state (F$_{100}$ = (1.30 $\pm$ 0.05) $\times$10$^{-6}$ ph cm$^{-2}$ s$^{-1}$, \fone\ = (-7.9 $\pm$ 0.2)$\times$10$^{-13}$ Hz s$^{-1}$) \citep{2018wang}, and the cameras were operated in the same modes.\\
Following the prescription of the 3XMM catalogue, for the pn camera we selected events with pattern 0 for energies from 0.2 to 0.5 keV, and 0--4 for energies between 0.5 and 10 keV \citep{2016marelli}. We extracted photons \ireviewmethree{within 20" of the pulsar position} determined by $Chandra,$ and we applied standard flags to remove bright columns.
These selections yielded samples of 1988 and 1896 X-rays for the first and second \xmm\ observation, respectively.\\ 
The \emph{Chandra} position was also used to barycenter the X-rays using the tool {\tt barycen} included in the XMM Science Analysis Software suite version 15.0\footnote{\url{https://www.cosmos.esa.int/web/xmm-newton/sas}} \ireviewmethree{and the JPL DE405} \ireviewmetwo{planetary ephemeris used by {\tt gtbary}}.

\subsection{Optical}


\jreviewme{\gamcyg\ was observed} with the Spanish 10.4 m Gran Telescopio Canarias (GTC) at the La Palma Observatory on  2016 June 8 (MJD 57547) under programme GTC27-16A (PI. N. Rea) 
in coincidence with the plateau of the \gam-ray flux after the rise observed 
in late 2014. The observations were performed 
with the camera called optical system for imaging and low-resolution integrated spectroscopy (OSIRIS)
equipped with a two-chip E2V CCD detector with a mosaic field--of--view of $7\farcm8 \times 7\farcm8$ and a pixel size of 0\farcs25 ($2\times2$ binning).
To remove cosmic-ray hits,  we took two sequences of 15 exposures each in the $g'$  ($\lambda=4815$ \AA; $\Delta \lambda=1530$\AA) and $r'$ ($\lambda=6410$ \AA; $\Delta \lambda=1760$\AA) filters  \citep{1996fukugita} with an exposure time of 145  s  to minimise saturation of bright stars. We chose the aim point in chip 2 with a 30\arcsec\ offset to the east to place the bright star BD+39 4152 ($V\sim 8.6$) in chip 1 and minimise 
scattered-light contamination.
The exposures were dithered by 10\arcsec\  steps in right ascension and declination, adequate to keep BD+39 4152 in chip 1 and our target at a safe distance from the CCD gap.
%
Observations were performed in dark time and under photometric  conditions, with a 0\farcs7 seeing and 1.1 airmass. 
Short (0.5--3 s) exposures of standard star fields \citep{Smith2002} were also acquired 
for photometry calibration
together with twilight sky flat fields.
We reduced our data  
using standard procedures in the {\sc IRAF} package {\sc ccdred}. We then aligned and co-added the single dithered exposures  using the task {\tt drizzle} to apply a cosmic-ray filter.

\section{\gam-\ and X-ray timing analysis}

To compare the X- and \gam-ray pulse profiles of \gamcyg\ during the two \xmm\ observations, we adopted two different timing analyses. The first analysis was based on a global timing solution obtained from \fermi-LAT photons and covered both \xmm\ epochs. Because the \gam-ray peaks are stable relative to this timing solution, it provides an absolute phase reference against which the phases of the \xmm\ pulse profiles can be measured.\\
To complement this timing analysis, we applied a second method that relies on a local \gam-ray timing solution built using \ireviewme{\fermi-LAT data during }a small time interval around the \xmm\ observations. In this case, the phase of the \gam-ray peaks might be different at the two epochs, \ireviewmethree{but we were still able to accurately measure the relative phase difference between the X-ray and \gam-ray peaks}.\\
We note that these methods work because both the \gam\ and X-ray phaseograms are based on the same reliable phase reference in order to compute reliable phase differences. For instance, the computation of the phases is handled by TEMPO2 \citep{2006hobbs} with the TZRMJD parameter defining phase zero at the MJD of the first time of arrival (ToA), \ireviewmetwo{ as well as TZRFRQ to define the frequency of the arrival time corresponding to TZRMJD and TZRSITE as the code of the reference site (e.g. the Solar System barycenter)}.
\jreviewme{By comparing the photon rotational phases computed by the {\tt fermi} and {\tt photon} TEMPO2 plugins as well as our scripts, we verified that all these tools handle the reference phase consistently.}

\subsection{\ireviewme{Full-mission timing analysis}}
\label{sec:absolute}
\begin{figure}[t]
\centering
\includegraphics[width=\hsize]{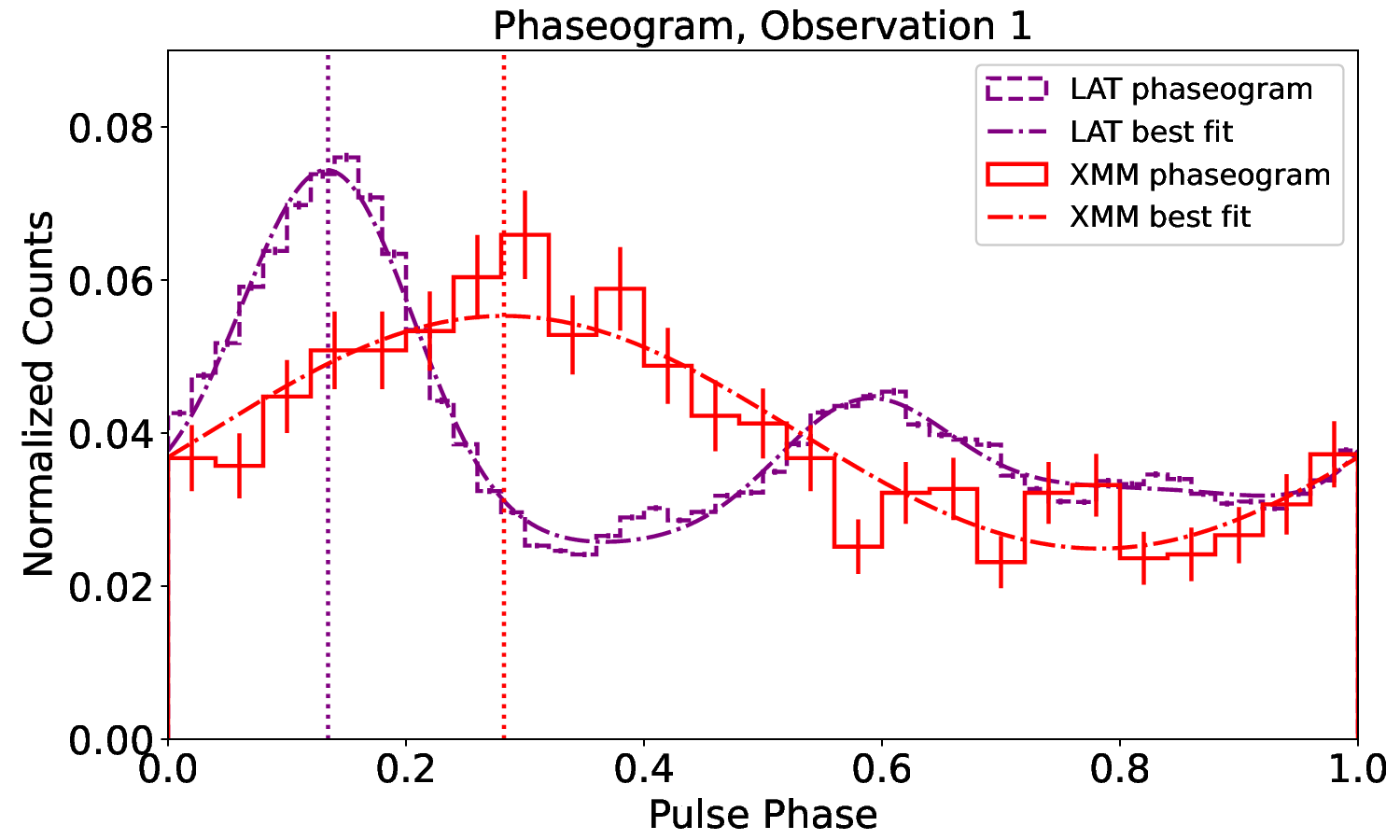}\\
\includegraphics[width=\hsize]{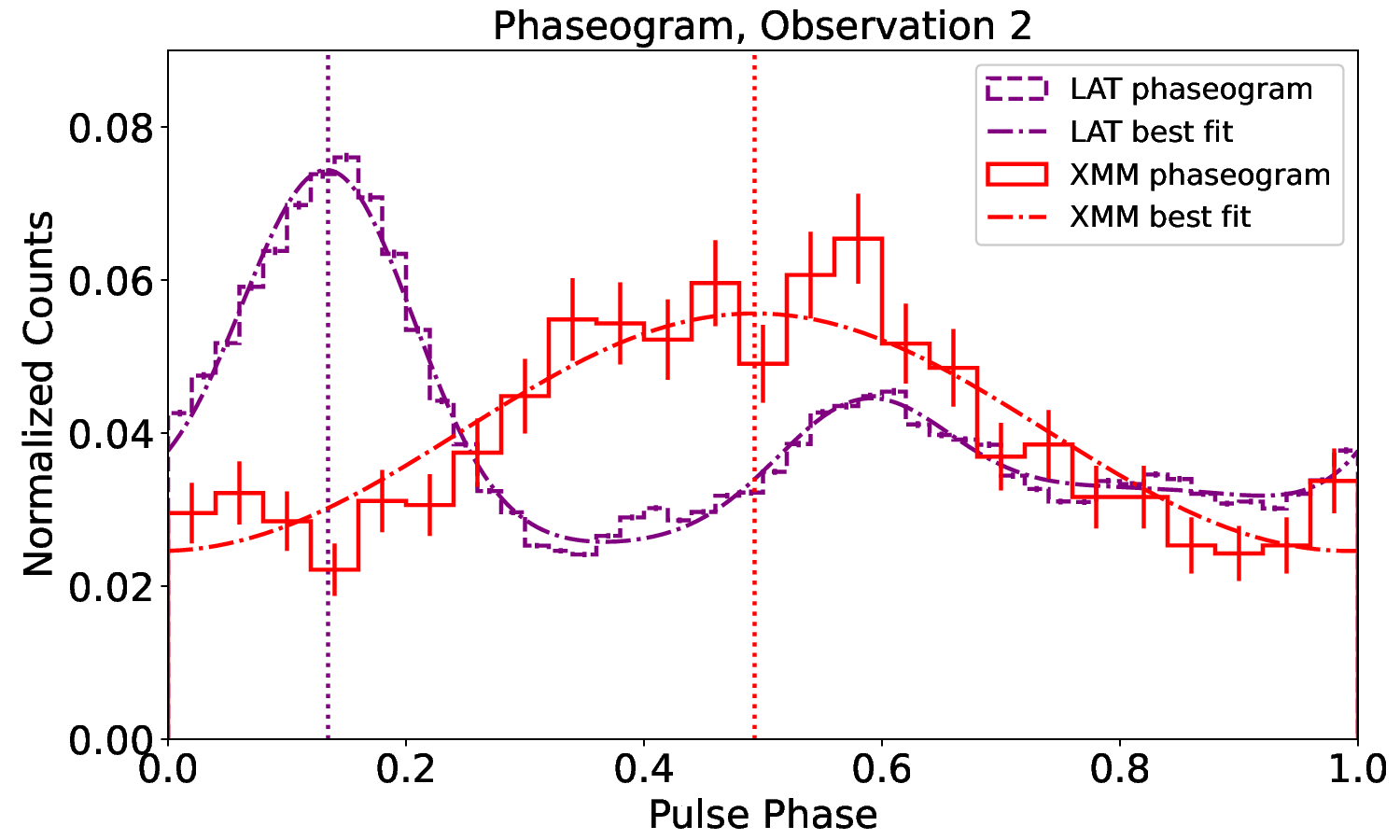}
\caption{Results from the full-mission timing analysis using best fits to the pulse profiles as described in Section \ref{sec:absolute}. \ireviewmethree{The top and bottom plots refer to the first and second \xmm\ observations, respectively. 
\fermi-LAT counts were calculated using photon weights. \xmm\ counts were normalized to 1, while \fermi-LAT counts were normalized to 2 for display purposes. \jreviewme{Counts from the diffuse \gam-ray background are not shown}. Statistical error bars are included for both phaseograms. The dotted vertical lines indicate the best-fit locations of the X-ray peak and higher \gam-ray peak.}}
\label{FigAbstiming}
\end{figure}

\ireviewme{For this first timing analysis}, we built a timing solution spanning the first $\sim$13 years of the \fermi-LAT mission. We employed an unbinned maximum likelihood method largely identical to the method of \citet{2022ajello}.  In brief, we simultaneously fitted the parameters of the timing model along with a stochastic model for timing noise. This timing noise model was implemented via a truncated Fourier series, and the amplitudes were constrained to follow a power law in frequency. The number of frequencies represented in the Fourier series (50) was chosen such that the highest frequency represented in the timing noise model produces a contribution that lay below the estimated white noise level in the data. The best-fit power-law index for the timing noise process is $\propto f^{-5.9}$. Consequently, there is little capability for the timing noise model to absorb high-frequency features, such as abrupt changes in phase or pulse shape. Additionally, we included the step changes in the spin-down rate \fone\ associated with state changes as parameters of the timing model. The resulting timing model accompanies this work in the supplemental material\footnote{\url{https://www-glast.stanford.edu/pub_data/1716/}}.\\
\ireviewme{Although small variations in the phaseograms during the first mode change were observed} \citep{2013allafort}\ireviewme{, the two main peaks did not change in phase. Therefore, the template used to build the TOAs is valid for the entire \fermi-LAT dataset under consideration.}
We used this timing solution to compute rotational phases of \gam-{} and X-rays using the {\sc fermi} and {\sc photons} plugins of TEMPO2 \citep{2006hobbs}, respectively. 
\ireviewme{During this part of the analysis, an important aspect of comparing phaseograms is the calculation of reference phase $\phi_{0}$. \ireviewmethree{To this end, we note that both plugins calculate $\phi_{0}$ in the same way, starting from the value of the TZRMJD parameter in the TEMPO2 timing solution PAR file.}}
The reference \gam-ray phaseogram was built using the full \fermi-LAT data set, which therefore covers both \xmm\ observations. \ireviewmethree{For each \jreviewme{\gam-ray photon,} a weight was computed using the {\tt gtsrcprob} tool, corresponding to the probability of having been emitted by the pulsar and the spectra of the pulsar and nearby sources as represented in the
4FGL-DR2 catalogue \citep{2020abdollahi}}. The \gam-ray timing solution covers the whole time range between the two \xmm\ observations, and it was built so that the two \gam-ray peaks of \gamcyg\ always occurred at the same phase \jreviewme{($\sim$0.15 for the first peak)} and act as a reference for computing the phase lag of the X-ray phaseograms (Figure  \ref{FigAbstiming}). \\
\ireviewmethree{A possible method for evaluating the phase lag between \gam-\ and X-ray pulsed emission is to compute the maximum correlation coefficient between the pulse profiles. However, this method yields a large uncertainty on the value of the correlation coefficient. In order to evaluate this uncertainty, we ran 1000 Monte Carlo realisations of the \xmm\ pulse profile, one with the number of photons of the first and one with those of the second observations. For each, we computed the correlation coefficients with respect to the LAT pulse profile, plotting and computing the mean and standard deviation. The resulting standard deviation corresponds to a 100\% uncertainty on the \gam-to-X relative phase, which is too large for the method to be considered robust enough.} \\
We therefore applied a second method based on fitting the pulse profile. \ireviewme{Using the PINT package \citep{pint}, we performed a maximum likelihood fit to the phaseograms obtained by assigning phases with the TEMPO2 plugins.} For this purpose, we collected \fermi-LAT data in a 10\degree\ region around the pulsar and chose a bin size of 0.02 in phase. We included \gam-ray weights, which
allowed us to estimate the \gam-ray background due to the diffuse emission, enhancing the significance of the peaks. \ireviewmefour{Having fewer counts from the \xmm\ observations, we were able to apply an unbinned method to fit the X-ray pulse profile. In Figure \ref{FigAbstiming} we use bins of size 0.04 to represent the X-ray phaseograms.} The \gam-ray profile was modelled using three Gaussian peaks, \ireviewmethree{while the X-ray profile required only one Gaussian peak. 
\jreviewme{For the \gam-ray peak, we also tested models with as many as four Gaussian peaks and combinations of asymmetric and symmetric Gaussians. The results did not change significantly, and we kept the simplest three Gaussian model. We also tested more elaborate models for the X-ray peaks, assessing the likelihood significance for a double Gaussian peak. For both the first and second observations, a second Gaussian component did not increase the significance beyond 5$\sigma$, and therefore we used a single Gaussian.}
Both models included a constant that represented the unpulsed component}. We measured the phase lags between the X-ray peaks in the two observations and the highest peak of the \gam-ray profile. We obtained $\Delta\phi_{\textup{peak}}$=0.147$\pm$0.013 in the first observation and $\Delta\phi_{\textup{peak}}$=0.358$\pm$0.014 in the second observation. The resulting phase delay between the two observations is 0.21$\pm$0.02.

\subsection{\ireviewme{Local timing analysis}}
\label{sec:relative}

    \begin{table}[t]
    \centering
      \caption{Parameters and results of the local timing test \ireviewme{for the two \xmm\ observations discussed in \citet{2013lin} (Obs. 1) and \citet{2018wang}(Obs. 2). For reference, we note that the \fermi-LAT mode change occurred at MJD 55850.}}
          \label{tab:table-reltiming}
          \begin{tabular}{lll}
             \hline
             \noalign{\smallskip}
                  &  Obs. 1 & Obs. 2 \\
             \noalign{\smallskip}
              \hline
              \hline
              \noalign{\smallskip}
            Epoch MJD       &  56029.1 & 57377.2 \\
              \noalign{\smallskip}
              \noalign{\smallskip}
             LAT Start MJD       &  56014.10 & 57354.7 \\
              \noalign{\smallskip}
              \noalign{\smallskip}
             LAT End MJD      &  56044.10 & 57399.7 \\
              \noalign{\smallskip}
              \noalign{\smallskip}
             LAT Duration       &  30 days & 45 days \\
              \noalign{\smallskip}
              \noalign{\smallskip}
             X-ray Photons       &  1385 & 1984 \\
              \noalign{\smallskip}
              \noalign{\smallskip}
             \fzero\ (Hz)    &  3.7689931$\pm 10^{-7}$ & 3.7688994$\pm 10^{-7}$ \\
              \noalign{\smallskip}
              \noalign{\smallskip}
             \fone\ (10$^{-13}$ Hz s$^{-1}$)      &  $-8.11\pm0.01$ & $-7.70\pm0.01$ \\
              \noalign{\smallskip}
              \noalign{\smallskip}
             $H$       &  74.3 & 45.2 \\
              \noalign{\smallskip}
              \noalign{\smallskip}
              $\Delta \phi _{\textup{peak}}$      &  $0.14\pm0.02$ & $0.35\pm0.02$ \\
              \noalign{\smallskip}
              \hline
          \end{tabular} 
 \end{table}

\begin{figure}[t]
\centering
\includegraphics[width=\hsize]{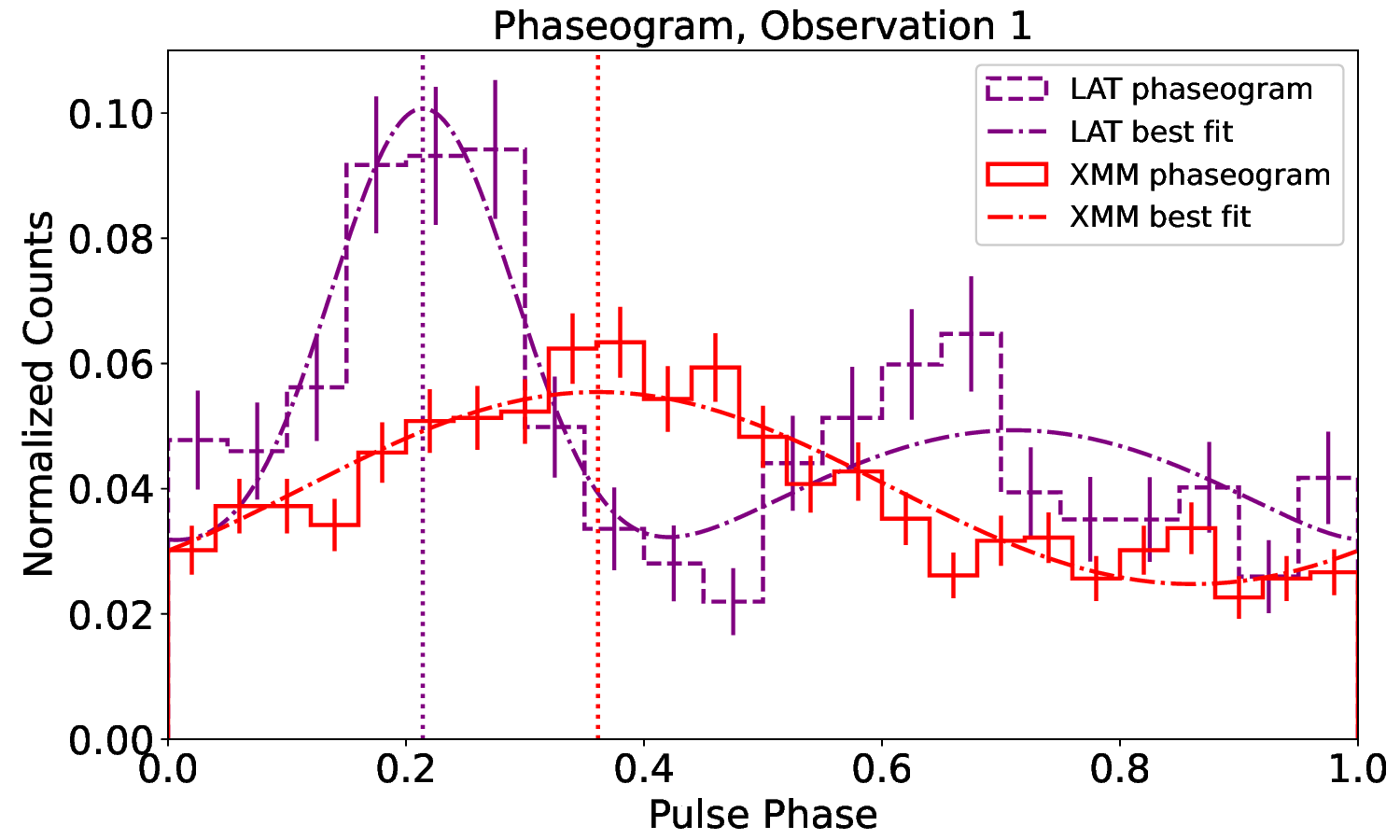} \\
\includegraphics[width=1\hsize]{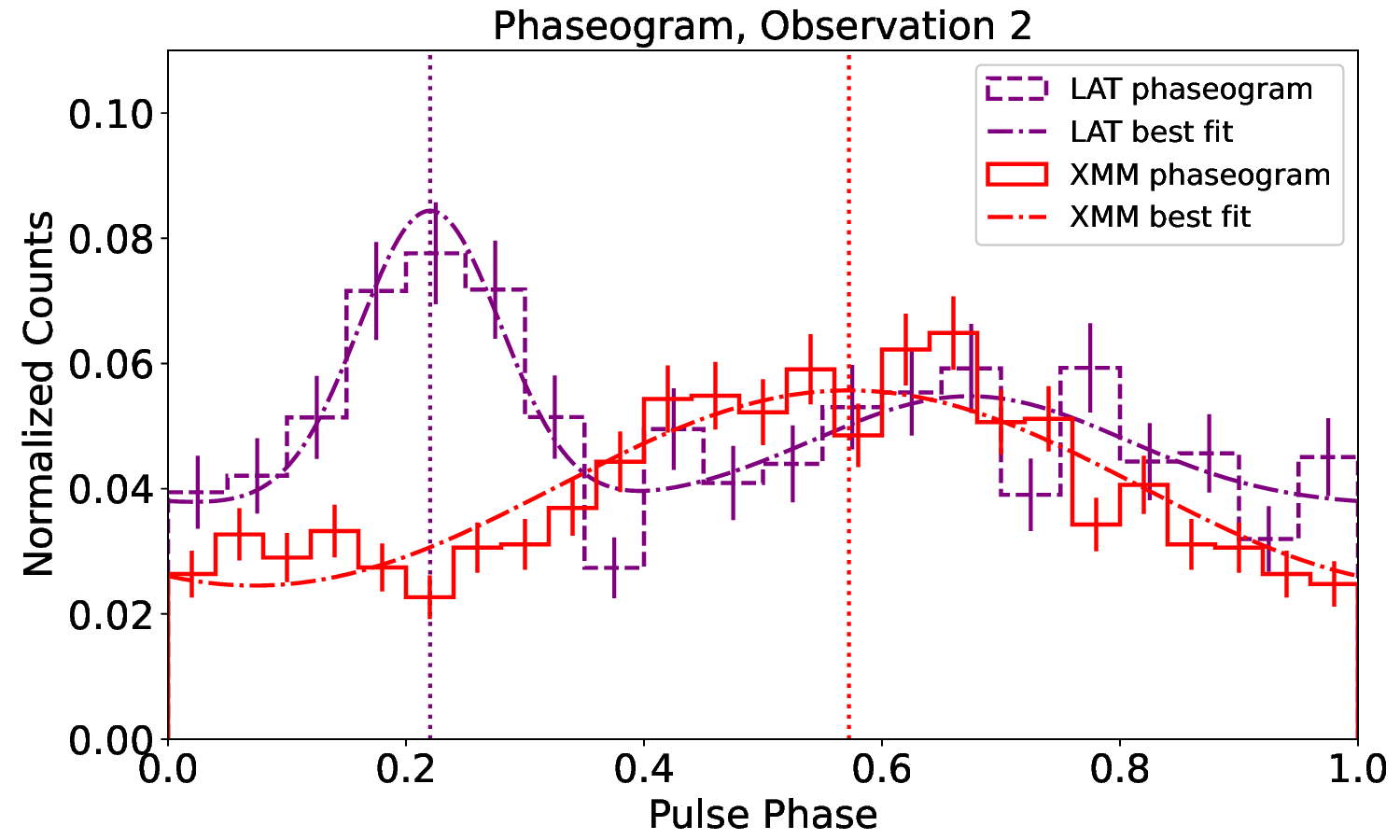}
\caption{
Results from the local timing analysis described in Section \ref{sec:relative}. \ireviewmethree{The top and bottom plots refer to the first and second \xmm\ observations, respectively. 
\fermi-LAT counts were calculated using photon weights. Both \fermi-LAT and \xmm\ counts were normalised to 1. \jreviewme{Counts from the diffuse \gam-ray background are not shown}. Statistical error bars are included for both phaseograms. The dotted vertical lines indicate the best-fit locations of the X-ray peaks and higher \gam-ray peaks. An initial phase $\phi_{0}$=-0.19 was introduced in the first observation mostly for display purposes in order to align the \gam-ray phaseogram with that of the second observation.}}
\label{fig:FigRelTiming}
\end{figure}

We also performed a test using a local timing solution built using a time window centred on each \xmm\ observation. The aim of the test was to avoid possible biases introduced by using the many parameters 
\ireviewme{of the global timing solution, as described at the beginning of Section \ref{sec:absolute}}.
\ireviewme{In these short windows, the time evolution of frequency can be adequately described using only the frequency and the first derivative. We thus selected \gam{} rays around each \xmm\ observation and searched for the pair of frequency, \fzero, and frequency derivative, \fone, that maximised the H-test statistics \citep{2010dejager}. 
\jreviewme{The search uses steps $df_{0}$ and $df_1$ related to the observation time $T$ as $df_0$=k/T and $df_1$=2k/T$^{2}$, where $k$ is an oversampling factor set to 0.3}
For this search over the \fzero-\fone\ plane, we used photons within 0\fdg9 of \gamcyg\ and with energies E>200 MeV, selection cuts that maximised the value of the H test.}
The length of the time window was chosen to provide a value of $H$ corresponding to a pulsation significance greater than 5$\sigma$, using the same event cuts reported in Section 2.1. As shown in Table \ref{tab:table-reltiming}, we have found that a 30-day window around the first \xmm\ observation (MDJ 56028) gives $H$=74.3. Around the second \xmm\ observation (MJD 57376), however, we required a 45-day window to produce a value $H$=45.2, suggesting a change in the pulsed fraction of the pulsar occurring between the two \xmm\ observations. This is consistent with the pulse profiles shown in Figure \ref{fig:FigRelTiming}, where the second peak is less evident at the second epoch, \ireviewme{ suggesting a change inthe  ratio of the heights of the first and second peak (P1/P2)  similar to that suggested by \citet{2013allafort}}. A comparison between the \gam-ray phaseograms around the X-ray observations shows that the \ireviewme{pulsed fraction changes from 35.6\% to 21.5\%}. This is also reflected in a change in the slope of the evolution of H-test versus time, as found using the full-interval \fermi-LAT timing solution.\\
Using the timing solutions reported in Table \ref{tab:table-reltiming}, we assigned rotational phases to the \xmm\ barycentered photons and evaluated the phase lag between \gam\ and X-ray profiles using the correlation function. We performed phase calculations 
using a custom script that computed the Taylor series expansion for the rotational phases,
\begin{equation}
    \phi(t) = \phi_{0} + f_{0}(t-t_{0})+f_{1}(t-t_{0})^{2}
\label{eq:taylor}
,\end{equation}
\ireviewme{where the epoch $t_{0}$ was set to the \xmm\ observation time. The reference phase $\phi_{0}$ was set to -0.19 for the first \xmm\ observation and to 0 for the second in order to align the two phaseograms. We note that this is mostly for display purposes because \ireviewmethree{the relevant measurement} is the phase difference between the X and \gam\ pulse profile in each observation, rather than the relative phaseogram alignment between observations. We also note that the term $\phi_{0}$ can be easily \ireviewmethree{converted into} a proper value of the TZRMJD TEMPO2 parameter.} \\
We repeated the maximum likelihood fit \jreviewme{and adopted the same phaseogram models} as in Section \ref{sec:absolute} using the \ireviewme{same two short time windows used to build the local timing solutions for this analysis}. In this step, we used bins of size 0.05 for the \gam\ phaseogram, 0.04 for the X-ray phaseogram. We obtained $\Delta\phi_{\textup{peak}}$=0.14$\pm$0.02 in the first observation and $\Delta\phi_{\textup{peak}}$=0.35$\pm$0.02 in the second observation, with a phase delay of 0.21$\pm$0.03. 
\ireviewmethree{The results, shown in Fig. \ref{fig:FigRelTiming} and summarised in Table \ref{tab:table-reltiming}, agree with the optimal parameters reported previously.} The timing solutions used for the analyses in Sections \ref{sec:absolute} and \ref{sec:relative} are provided in the supplementary material \jreviewme{as well as in the Fermi-LAT publication board website\footnote{\url{https://www-glast.stanford.edu/pub_data/1716}}}.

\subsection{Further timing checks}
\begin{figure}
\centering
\includegraphics[width=\hsize]{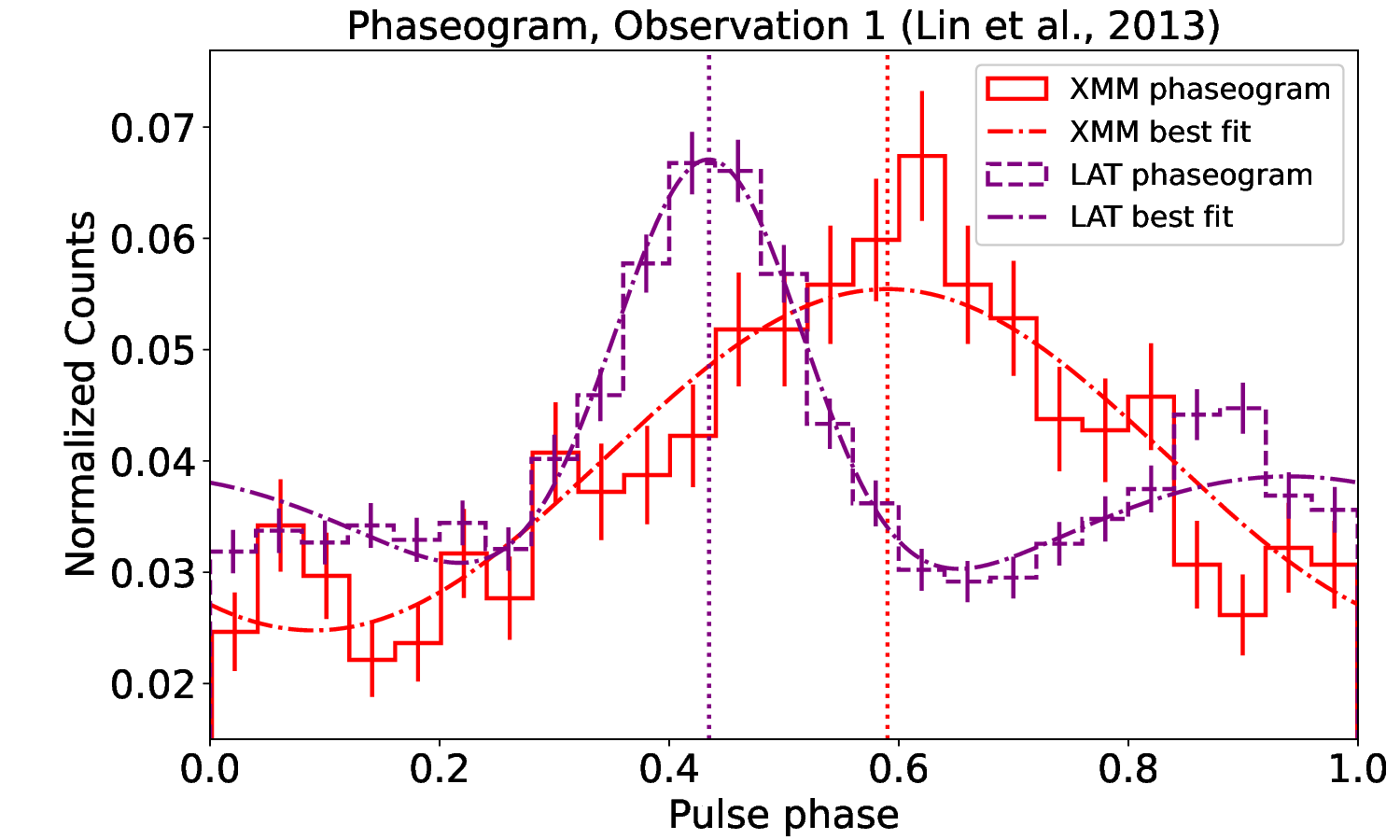}
\includegraphics[width=\hsize]{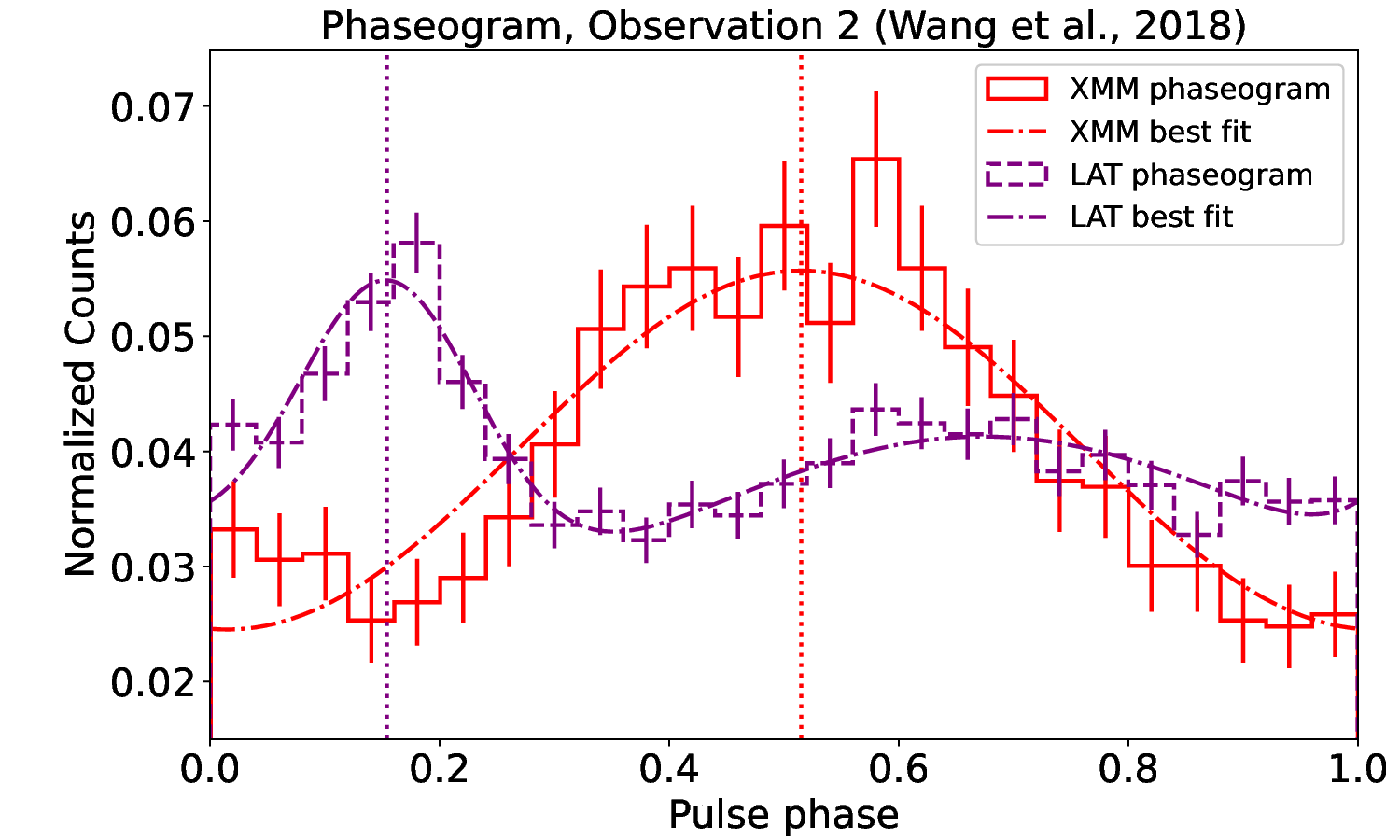}
\caption{\ireviewmefour{Results of the timing checks that were performed based on the literature. The pulse profiles were obtained using the timing solutions published in \citet{2013lin} (\emph{top}) and \citet{2018wang} (\emph{bottom}). Both \fermi-LAT and \xmm\ counts were normalised to 1. Statistical error bars are included for both phaseograms. The dotted vertical lines indicate the best-fit locations of the X-ray peak and higher \gam-ray peak. No photon weights were applied.}}
\label{fig:FigLiteratureCheck}
\end{figure}

By comparing the profiles using these two timing analyses, we found a consistent change in the phase lag between \gam-\ and X-rays. Because this substantially disagrees with the results of \citet{2018wang}, we performed a series of additional tests to confirm the validity of this analysis. \\
\citet{2018wang} concluded that the phase lag between X and \gam\ phaseogram is $\sim$0.14 for both the first and second \xmm\ observation. To determine the reason for the discrepancy with our results, we tried to reproduce the pulse profiles reported in \citet{2013lin} and \citet{2018wang}. To do this, we used the timing solution for the first \xmm\ observation reported in \citet{2013lin} and that reported in \citet{2018wang} for the second observation and calculated the phases directly by using the Taylor series expansion for the rotational phase (Equation \ref{eq:taylor}). \ireviewmefour{We fit models as described in Section \ref{sec:absolute} and \ref{sec:relative} to determine the relative phases of the \gam-\ and X-ray pulse profiles.} The results are reported in Figure \ref{fig:FigLiteratureCheck}.\\
\ireviewmefour{The profile obtained for the first \xmm\ observation using the timing solution in \citet{2013lin} shows a phase lag of 0.155$\pm$0.014, which is consistent with the finding of \citet{2018wang} (cf. 0.14), as well as with the phase lag of 0.150$\pm$0.016 obtained from our analysis.}
\ireviewmefour{However, while the pulse profiles in \citet{2018wang} indicate a $\sim$0.14 phase lag, the profiles from our re-analysis indicate a phase lag of 
0.361$\pm$0.015, which appears to be inconsistent with the finding of \citet{2018wang} and in agreement with our full-mission and local timing analyses.} 
\ireviewmethree{A possible explanation could be related to the fact that the method based on correlations, also used by \citet{2018wang}, provides extremely large uncertainties, as shown in Section \ref{sec:absolute}, and therefore cannot be considered robust enough to draw conclusions.}\\
\ireviewme{\ireviewmethree{We also considered whether the change in phase lag that we observed could be related} to drifts in the \fermi-LAT clock that might cause different \gam-to-X relative phase at different epochs. However, \citet{2021ajello} have shown that the \fermi-LAT \ireviewmethree{time stamp for each photon is} accurate to 300 ns in absolute time.}\\ 
Finally, we addressed whether the discrepancy in phase measurements could be related to the timing accuracy in \xmm\ \citep{2012martin}. We searched for anomalous time jumps in all pn small-window observations of \gamcyg,\ but found none, and therefore, we exclude \xmm\ timing jumps as the cause of the phase shift.\\
As a final check, we performed our local timing analysis on the Crab pulsar, which has been used to evaluate the \xmm\ timing accuracy \citep{2012martin}. We report the details of this analysis in Appendix A. Our conclusion is that there is no change in phase lag between the \fermi-LAT and \xmm\ pulses of the Crab, which confirms that the results reported in this section are unlikely to be an artefact of our analysis methods.

\subsection{Multi-epoch X-ray spectroscopy} \label{sec:xspec}

We performed a standard spectral analysis of the X-ray data of \gamcyg\ at the two different epochs of the {\it XMM-Newton} observations: the first analysis was made $\text{about }$6 months after the first mode change, and the second one  was made when the $\gamma$-ray flux had recovered to the initial state (see Table \ref{tab:table-reltiming}).
We performed a standard search for high particle background periods \citep[following][]{2005deluca} and excluded them, thus obtaining good exposure times of $\sim$80 ks and $\sim$95 ks for the two observations, respectively. We extracted photons within 30\arcsec\ \ireviewmethree{of the pulsar position}, as determined by {\it Chandra}, to obtain the source spectra. Background spectra were extracted from nearby source-free regions. We extracted 0--4 pattern events for pn and 0--12 for the MOS detectors. Finally, we grouped spectra to have at least 25 counts per bin\jreviewme{, resulting in 16 energy bins in the 0.3-12 keV range.} \\
\begin{figure*}
\centering
\includegraphics[width=\hsize]{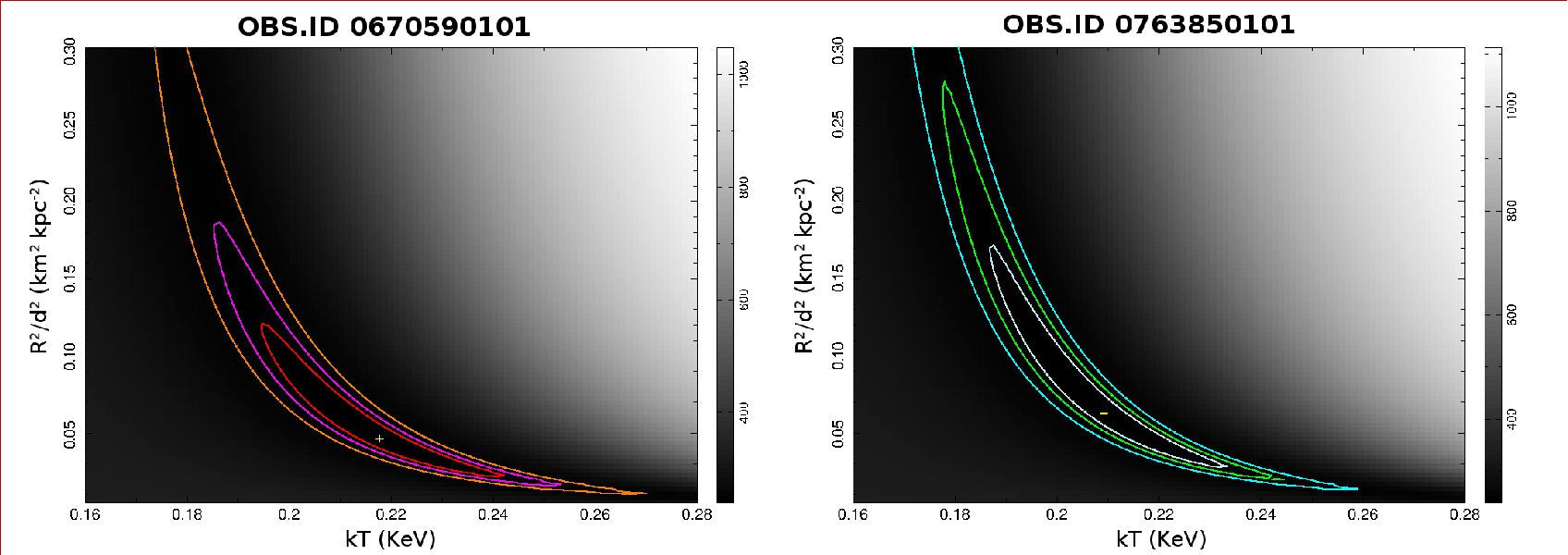}
\caption{Contour plots of the thermal model parameters for the two different {\it XMM-Newton} observations that were obtained as described in Section \ref{sec:xspec}. We plot the temperature against the emitting radius. \ireviewmethree{We also report} the $\chi^2$ value for each realization, where the degrees of freedom are 208. We show the 1$\sigma$, 90\%, and 3$\sigma$ confidence level contours around the best fit (marked with a cross).}
\label{fig:FigXcont}
\end{figure*}
To fit the spectra, we used the {\tt XSPEC} package v.12.10.1 \citep{1996arnaud}, considering events in the 0.3--10 keV energy range.
We fitted the spectra of the two observations separately using either an absorbed blackbody ({\tt bbodyrad}) 
\citep[abundances from][XSPEC model {\tt tbabs}]{2000wilms} or an absorbed power law ({\tt powerlaw}). Single-component models give statistically unacceptable fits, with null hypothesis probabilities (n.h.p.) $<5\times10^{-5}$. 
Thus, we tried a two-component model (absorbed power law plus blackbody). For both observations, we obtained an acceptable fit with n.h.p. of 0.04 and 0.34, respectively. With the model settled, we searched for significant differences in the spectra and/or fluxes between the two observations.
We found no evidence of spectral shape or flux changes. The contemporaneous fits of both spectra with all the model parameters yielded results with acceptable n.h.p. of 0.067 (reduced $\chi^2$ of 1.15 with 210 degrees of freedom).
Using this model, we obtained a hydrogen absorption column density $N_{\rm H}$=(1.2$\pm$0.2)$\times10^{22}$ cm$^{-2}$, a blackbody temperature $kT$=(215$\pm$15) eV (consistent with \citet{2013lin}), a blackbody emitting radius of R=340$_{-80}^{+120}$ $d_{\rm 1.5}^2$ m (where $d_{\rm 1.5}$ is the distance in units of 1.5 kpc), and a photon index $\Gamma$=1.0$\pm$0.3 (all the errors are at 1$\sigma$ confidence). The unabsorbed thermal flux in the 0.3--12 keV band is (11.0$\pm$2.4)$\times10^{-14}$ erg cm$^{-2}$ s$^{-1}$ , and the unabsorbed non-thermal flux is (3.2$\pm$0.7)$\times10^{-14}$ erg cm$^{-2}$ s$^{-1}$. \\
Finally, we searched for changes in the parameters of the thermal component. We performed fits using the same model as above, but with the parameters of the thermal component left free to vary between the different observations. We obtained the contour plots reported in Figure \ref{fig:FigXcont} for the pairs of parameters describing the thermal emission of the two different observations, indicating consistent temperatures.

\section{Searching for an optical counterpart}
\begin{figure}[h]
\centering
\includegraphics[width=\hsize]{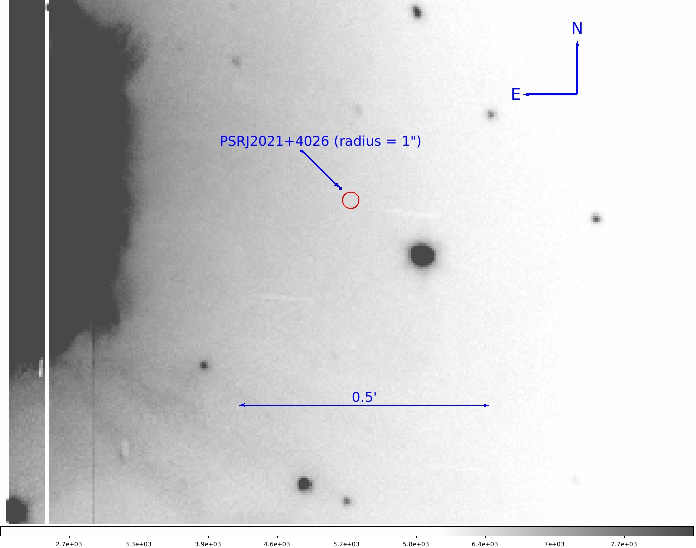}
\caption{GTC/OSIRIS zoomed $g'$-band image of the \gamcyg\ field. The pulsar position is marked by the red circle, whose radius accounts both for the absolute uncertainty on the pulsar {\em Chandra} coordinates and the accuracy of our astrometric solution. The intensity levels in the bottom bar are in ADU and  have been adjusted for a better clarity. The point-spread function wings of the bright star BD+39 4152 are visible on the left, overflowing from the gap between the two CCDs. The white band on the left corresponds to the CCD gap residual after image dithering.}
\label{fig:FigOptical}
\end{figure}

\begin{figure}
\centering
\includegraphics[width=\hsize]{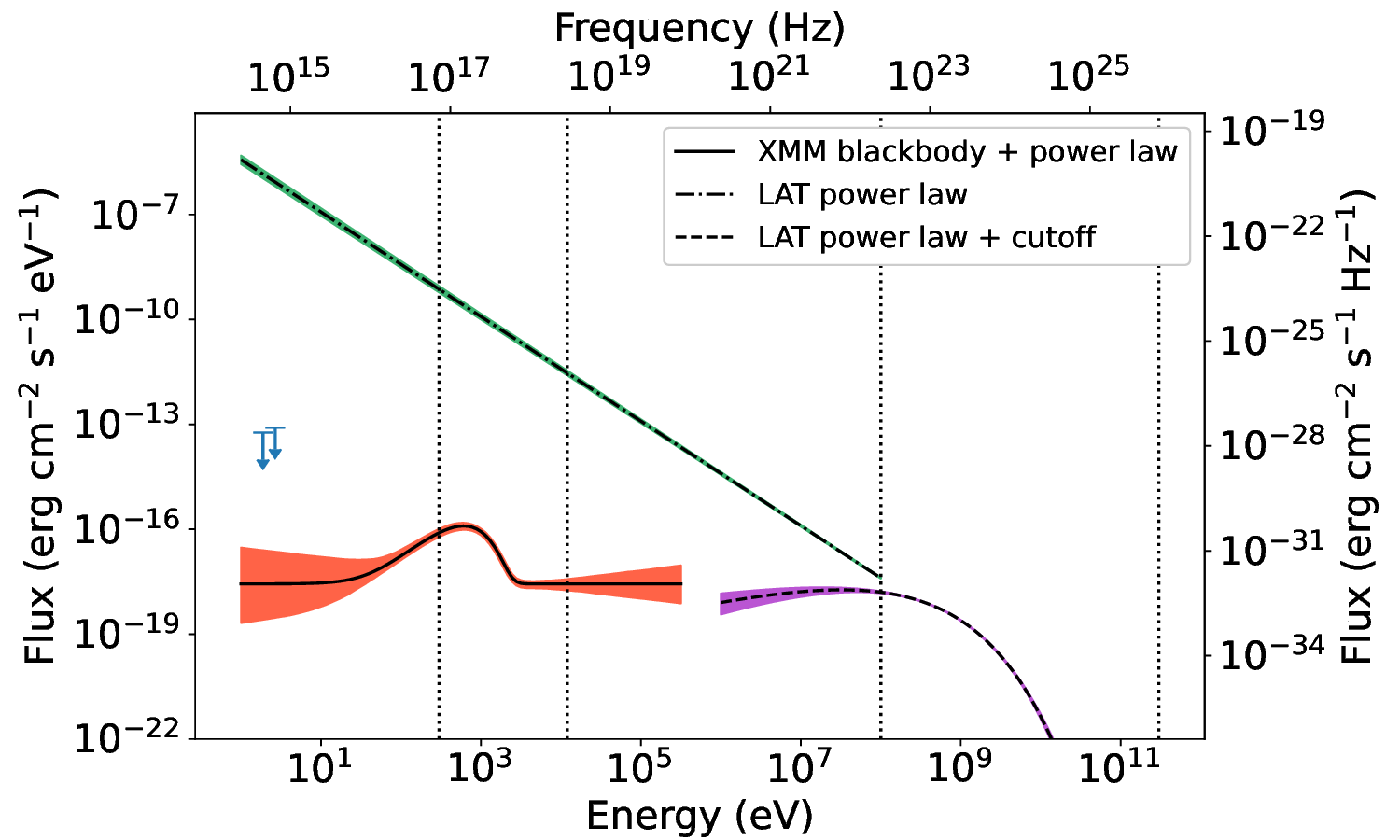}
\caption{\jreviewme{Best-fit functions and 1$\sigma$ confidence contours for the XMM and LAT spectra of \gamcyg. The parameters for the XMM spectral models are those reported in Section \ref{sec:xspec}. LAT spectral models are taken from the 4FGL-DR2 catalogue. Power laws were extrapolated down to 1 eV to highlight the observed discontinuity in the high-energy spectral index. The optical upper limits are reported in blue. The dotted vertical lines indicate the validity energy bands of the X-ray (0.3-12 keV) and \gam-ray (0.1 - 300 GeV) best-fit functions.}}
\label{fig:FigSpectra}
\end{figure}

We searched for the \gamcyg\ optical counterpart in the GTC/OSIRIS images using the {\em Chandra} coordinates reported in \cite{2011weisskopf}
as a reference, with a 99\% confidence level uncertainty of  
1\arcsec.
Since \gamcyg\ is radio quiet and no multi-epoch {\em Chandra}  observations 
are available, we could not measure its proper motion from {\em Chandra} observations, \ireviewme{and the whitening parameters required for this young and energetic pulsar \ireviewmethree{make it challenging to measure them directly from} \fermi-LAT timing.} For the distance of the $\gamma$ Cygni SNR ($1.5\pm0.4$ kpc), as determined by  \cite{1980landecker} and the average transverse pulsar velocity (400 km s$^{-1}$) by \cite{Hobbs2004}, we would expect an angular displacement of $\sim 0\farcs3$ in an unknown direction at the epoch of our GTC observations (MJD=57547),
which we must account for together with the nominal error on the pulsar position. 
%
We computed the astrometric solution on the GTC/OSIRIS images using the {\tt wcstools}\footnote{{\texttt http://tdc-www.harvard.edu/wcstools}}  software package, matching the sky and pixel coordinates of stars selected from the Two Micron All Sky Survey (2MASS) All-Sky Catalog of Point Sources \citep{Skrutskie2006}, 
with an overall accuracy 
of $\sim$0\farcs2.\\
Fig. \ref{fig:FigOptical} shows the OSIRIS co-added $g'$-band image of the \gamcyg\ field.
\ireviewme{Although star BD+39 4152 was offset to chip 1, its scattered light is still strong enough to increase the sky background at the aim point in Chip 2 in the $g'-$ and $r'$ -band images by 30\% and 35\%, respectively. We computed the increase in sky background by sampling the counts and rms at the pulsar position in a box of $25\times25$ pixels ($6\farcs25 \times 6\farcs25$), and, in the same way, in different regions at the western part of the CCD chip at the greatest distance from BD+39 4152, where the sky background is least intense, and averaged the measurements.} 
We computed the $3 \sigma$ limiting magnitudes at the pulsar position from the rms of the sky background \citep{Newberry1991}
,
obtaining $r' = 25.3$ and $g' = 26.1$, the deepest constraints to date on the pulsar brightness. 
We corrected these values for the interstellar reddening in  the pulsar direction, $E(B-V) =2.15 \pm 0.04$, derived from the $N_{\rm H}$
best-fitting the  {\em XMM-Newton} X-ray spectra (see previous section)
and the relation of \cite{PS1995}.
\jreviewme{We checked that this value is consistent with the Galactic extinction maps within 5’ of the location of the pulsar.}
For the estimated reddening value, the extinction-corrected flux upper limits 
at the  $g'$- and $r'$-band peak frequencies
are  $F_{g'}$ = $3.3 \times 10^{-28}$ and $F_{r'}$ = 2.4 $\times 10^{-28}$ erg cm$^{-2}$ s$^{-1}$ Hz$^{-1}$. 
\jreviewme{Fig. \ref{fig:FigSpectra} shows that the optical upper limits}
are below and above the extrapolations of the $\gamma-$ and X-ray power-law spectral models, respectively, underlining the discontinuity in the multiwavelength spectrum, a characteristic seen in other pulsars \citep{Mignani2018}.
Our $g'$-band upper limit corresponds to an unabsorbed flux $F_{g'} \la 6.8 \times 10^{-14}$ erg cm$^{-2}$ s$^{-1}$, implying an optical luminosity $L_{g'} \la 18.3 \times 10^{30}$ erg s$^{-1}$ $d^{2}_{1.5}$, consistent with the values observed in pulsars of the same age as \gamcyg\ \citep{Mignani2016}.

\section{Discussion}
\ireviewmethree{Our multiwavelength timing monitoring observations have revealed} a shift in the relative phase of the X- and \gam-ray pulses between two epochs having different \fone\ and \gam-ray flux. We can use the results based on \ireviewme{full mission and local timing analyses} reported in Sections \ref{sec:absolute} and \ref{sec:relative} to compute a weighted average of the phase lag, obtaining \ireviewmethree{0.21$\pm$0.02}. This shift could be related to the state changes undergone by the pulsar, and we present here a toy model to interpret these results.\\
The X-ray spectral analysis suggests that \ireviewmethree{the X-ray emission} could be due to polar cap heating. The blackbody + power-law fit presented in section \ref{sec:xspec} points to a thermal emission region of radius $\sim$ 340 m and a luminosity of $\sim$2.5$\times$10$^{31}$ erg s$^{-1}$, which matches the predicted polar cap heating luminosity of 10$^{31}$ erg s$^{-1}$ reported in \citet{2001harding} reasonably well. The measured  temperature of the polar cap is higher than we would expect for a cooling neutron star with $\sim$77 kyr such as \gamcyg\ \citep{2021Rigoselli,2020potekhin}. \\
The shift of the X-ray pulse phase by 0.21, if corresponding to the emission coming from one of the magnetic poles, is much larger than the size of a standard polar cap \citep{2004page}. \ireviewme{Here we speculate that the pulsar has a multipolar magnetic field, where the quadrupole component is dominant near the stellar surface.} However, \ireviewmethree{at large distances} from the neutron star, this quadrupole component is not easily detectable because the dipole field will dominate and the global magnetospheric structure will be that of a dipole, with the current sheet outside the light cylinder producing the emission observed at \gam-ray energies by \fermi\  \citep{kala2014}. \ireviewme{In this configuration, the high-altitude dipole-dominated field lines will need to connect with the poles of the quadrupole-dominated field lines, depending on the strength and alignment of this component. This connection would lead to a heating of some but not all of the poles. If an event, such as the mode change observed in the pulsar, happens to change either the surface multipolar field or the configuration of the current sheet, the current sheet could connect to a different pole. We expect that the poles of a star-centred quadrupole component are about 0.25 in phase apart; therefore, this reconfiguration might explain the observed phase shift in the X-ray peak. A multipolar field has in fact been strongly inferred for the millisecond pulsar PSR J0030+0451 from fitting of its thermal X-ray profile using NICER data \citep{kala2021}.  Quadrupole field components can produce polar caps that are larger and more extended that those of a pure dipole field \citep{2017Gralla}. \cite{2021Rigoselli}, using a more complex magnetic atmosphere model to fit the X-ray profile of \gamcyg, find a heated area much larger than dipole.} \\
This model could also explain the observed variability in \gam{} -rays. If the pole connected to the current sheet immediately after the first state change had properties more favourable to pair production (e.g. higher surface field strength, smaller field radius of the curvature), then an increase in pair multiplicity, or equivalently, in conductivity, would increase \fzero\ \citep{2012kala,2012li}. An increase in conductivity would also decrease the accelerating electric field in the current sheet because more pairs would result in greater screening, which would decrease the \gam-ray flux as observed during the mode changes in 2011 and 2014. Moreover, a decreased electric field would also lead to a lower cutoff energy.  \ireviewme{A decrease in the spectral cutoff energy was observed during the first mode change in 2011 \citep{2013allafort}.}\\
\jreviewme{The number of \gam-ray pulsars detected by the LAT has increased to nearly 300 since its launch\footnote{Public list of LAT detected pulsars available at: \url{https://confluence.slac.stanford.edu/display/GLAMCOG/Public+List+of+LAT-Detected+Gamma-Ray+Pulsars}}, however, PSR J2021+4026 is still the only isolated pulsar to have shown such evident mode changes. 
This unique behaviour is thus difficult to explain, but a combination of magnetospheric properties and viewing geometry may contribute.}
\ireviewme{\citet{2016Ng} suggest that the mode changes are associated with a reconfiguration of the magnetic field on the neutron star surface due to crustal plate tectonics.  Alternatively, quasi-periodic changes in the current sheet, due to changes in plasma supply to the magnetosphere \citep{2012bLi}, could cause the dipole field component to connect to different poles of the quadrupole component.}

\section{Conclusions}
We have presented an updated multiwavelength analysis of \gamcyg\ emission using \fermi-LAT, \xmm,\ and GTC data to better understand the physics behind the mode changes experienced by this pulsar in 2011 and 2014.
Through a simultaneous analysis of \xmm\ and \fermi{} data, we detected a shift in the relative phase between the X and \gam-ray pulses. This phase shift was not observed before, and so we have performed two complementary analyses using \ireviewme{a full mission, whitened timing solution (Section \ref{sec:absolute}) and two simpler timing solutions based on short time windows \ireviewmethree{centred around} the two \xmm\ observations (Section \ref{sec:relative})}.
Both analyses confirmed the phase shift, and they produce a consistent value of 0.21 \ireviewmetwo{$\pm$ 0.02} in phase.\\ 
Since the X-ray emission is consistent with that of a heated polar cap, we interpret this phase shift as a change in the connection of dipole field lines with inner quadrupole field components close to the stellar surface. We suggest that mode changes would affect the accelerating electric field, thus producing a simultaneous decrease in flux and increase in \fone, as observed for the pulsar.\\ 
Furthermore, we observed \gamcyg\  with the 10.4 m GTC to search for an optical counterpart whose flux was possibly enhanced following the mode change, but we could not detect it down to $g'$ =  26.1 and $r'$ = 25.3.
The background from star BD+39 4152 ($V\sim 8.6$) at $\sim 1\arcmin$ prevented the limits from being deeper.
\ireviewmethree{{\em } Observations with the Hubble Space Telescope would provide} the only possibility to detect the pulsar in the optical and pave the way to coordinated multiwavelength monitoring programmes.\\
The repeating nature of the variability event observed for \gamcyg\ on timescales of a few years suggests that the pulsar magnetosphere is affected by some periodic reconfiguration. A continuous monitoring of the source would be fundamental for understanding the physics of this unique \gam-ray-emitting neutron star in more detail.
\begin{acknowledgements}
The authors would like to thank the $Fermi$-LAT internal referee David Smith and the anonymous reviewer at the journal for their useful review and discussions which improved the manuscript.

The \textit{Fermi} LAT Collaboration acknowledges generous ongoing support
from a number of agencies and institutes that have supported both the
development and the operation of the LAT as well as scientific data analysis.
These include the National Aeronautics and Space Administration and the
Department of Energy in the United States, the Commissariat \`a l'Energie Atomique
and the Centre National de la Recherche Scientifique / Institut National de Physique
Nucl\'eaire et de Physique des Particules in France, the Agenzia Spaziale Italiana
and the Istituto Nazionale di Fisica Nucleare in Italy, the Ministry of Education,
Culture, Sports, Science and Technology (MEXT), High Energy Accelerator Research
Organization (KEK) and Japan Aerospace Exploration Agency (JAXA) in Japan, and
the K.~A.~Wallenberg Foundation, the Swedish Research Council and the
Swedish National Space Board in Sweden.
Additional support for science analysis during the operations phase is gratefully
acknowledged from the Istituto Nazionale di Astrofisica in Italy and the Centre
National d'\'Etudes Spatiales in France. This work performed in part under DOE
Contract DE-AC02-76SF00515.

Based on observations made with the Gran Telescopio Canarias (GTC), installed in the Spanish Observatorio del Roque de los Muchachos of the Instituto de Astrofísica de Canarias, in the island of La Palma.

IRAF is distributed by the National Optical 
Astronomy Observatories, which are operated by the Association of Universities for Research in Astronomy, Inc., under cooperative agreement with the National Science Foundation.

Work at NRL is supported by NASA.

\end{acknowledgements}

%
\bibliographystyle{aa} 
\bibliography{j2021ref} 
%
\begin{appendix} 
 \section{Timing checks using the Crab pulsar}
\begin{figure}
\centering
\includegraphics[width=\hsize]{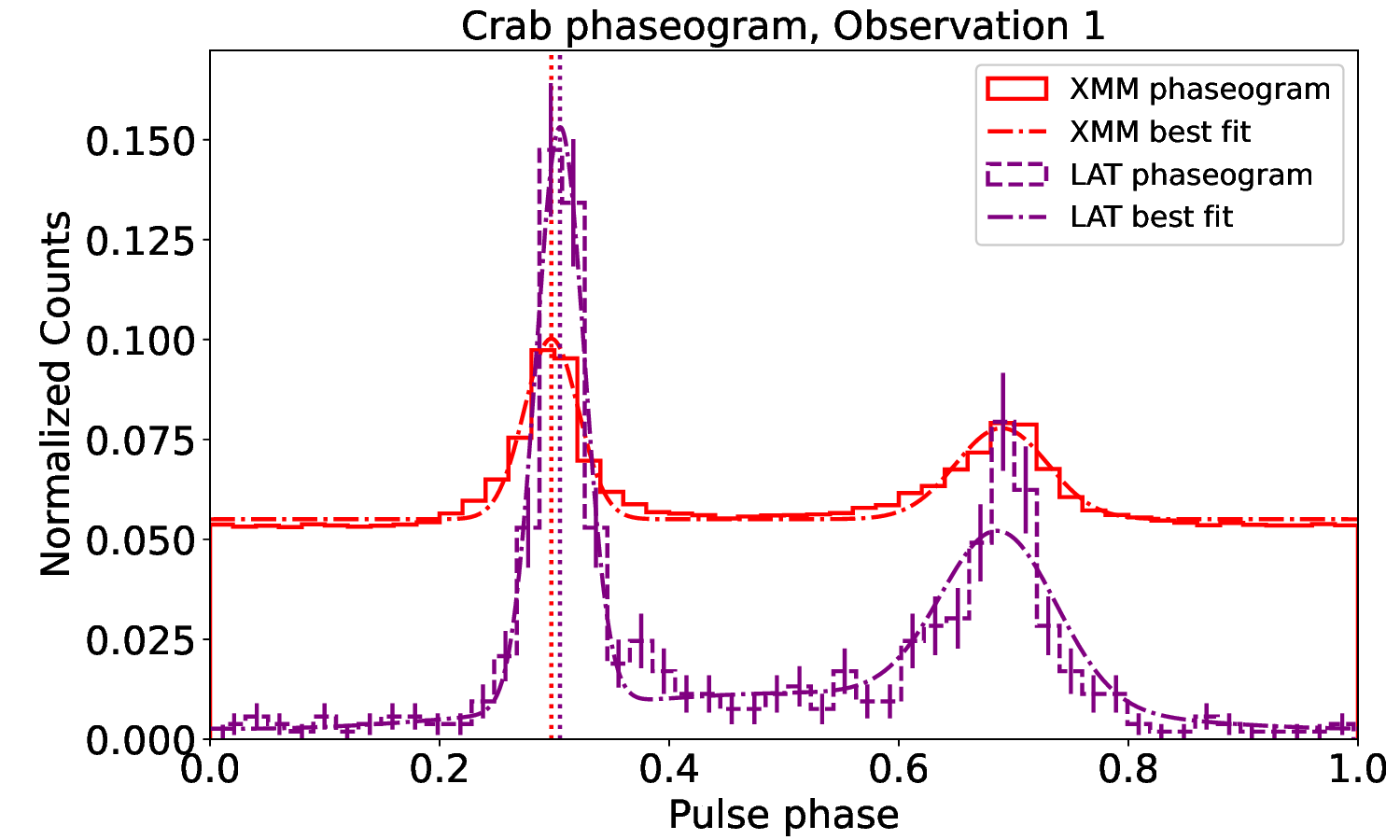}
\includegraphics[width=\hsize]{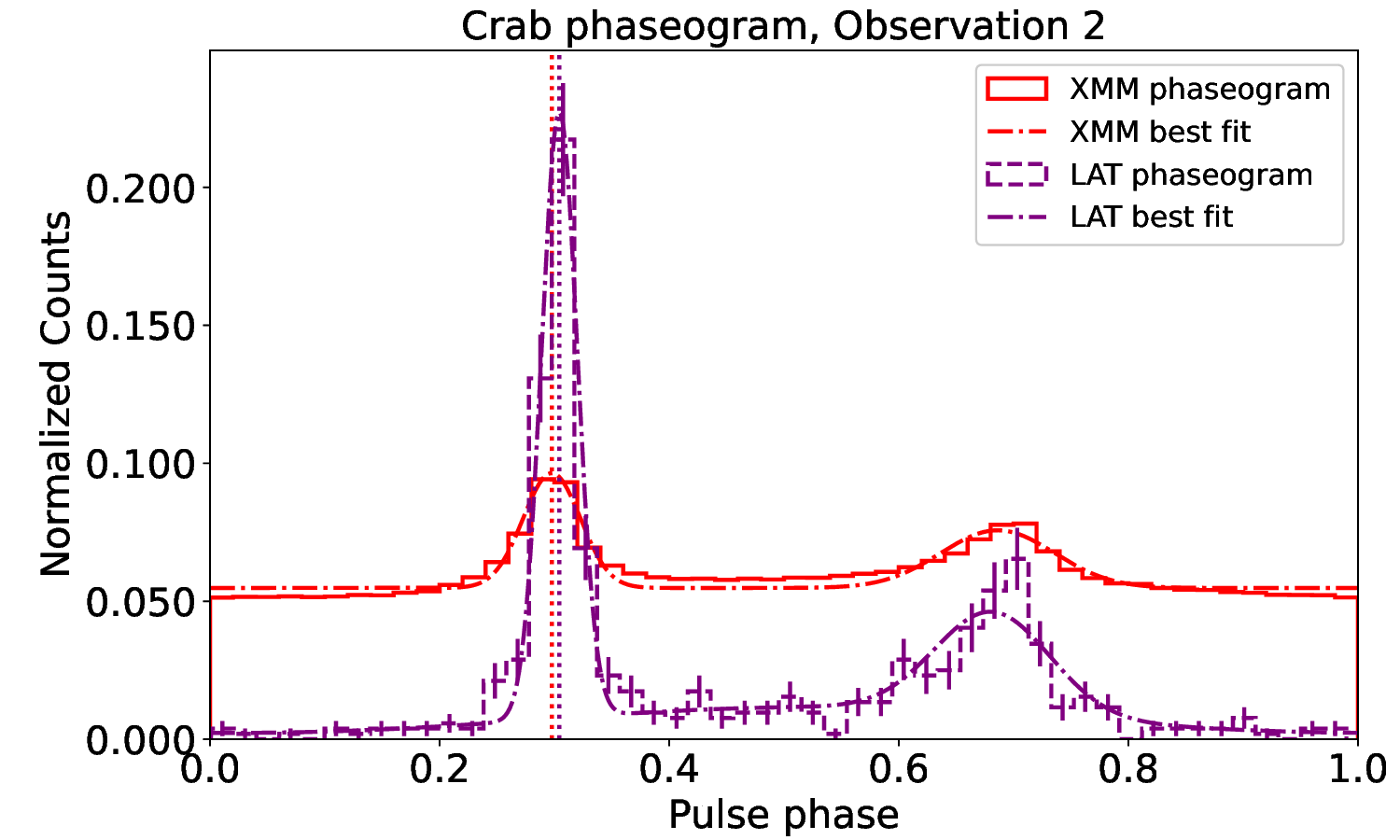}
\includegraphics[width=\hsize]{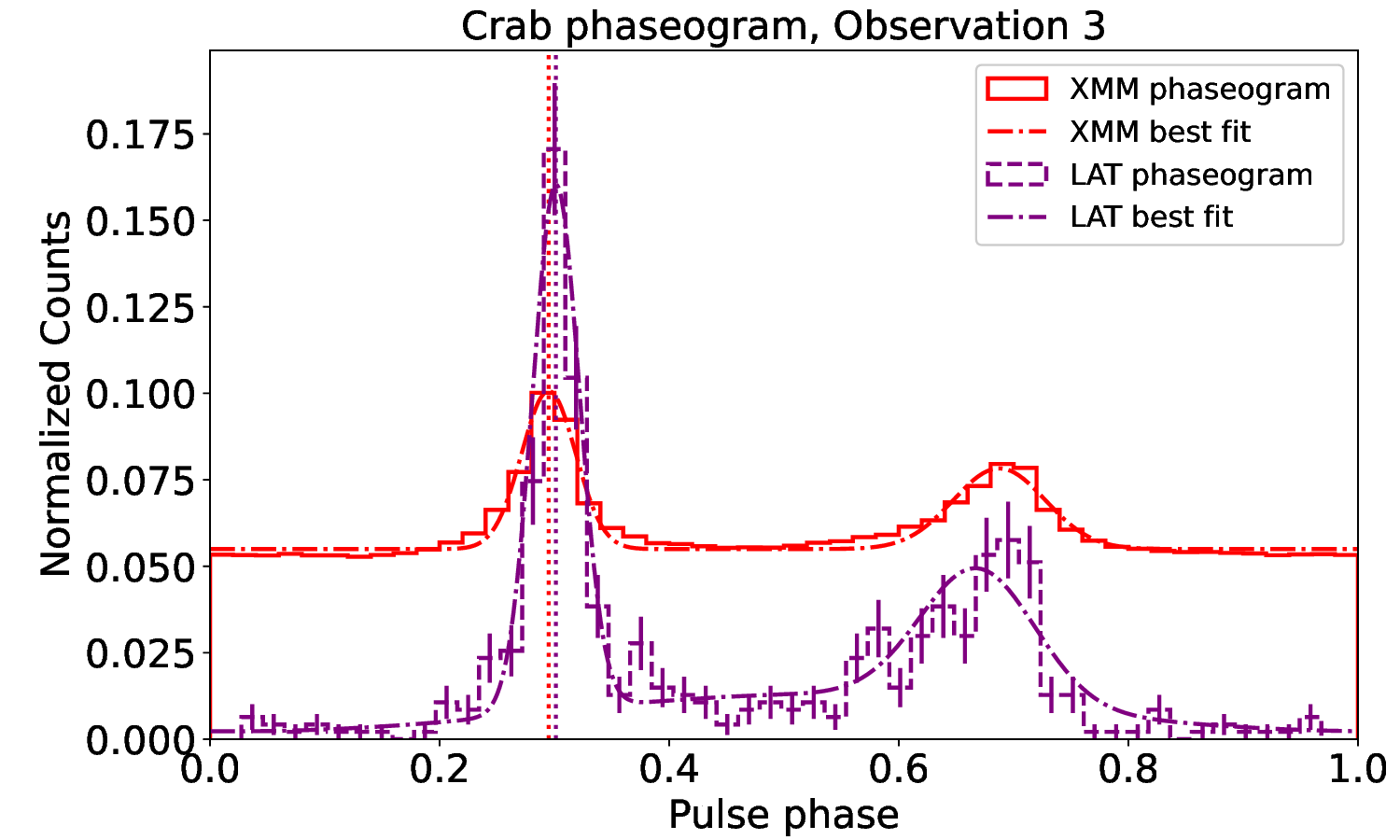}
\caption{\ireviewmefour{Results of the timing check made using three \xmm\ observations of the Crab pulsar. \fermi\ counts were normalized to 1, and \xmm\ counts were normalized to 3 for display purposes. Statistical error bars are included for both phaseograms. The potted vertical lines indicate the best-fit locations of
the X-ray peak and higher \gam-ray peak. No photon weights were applied.}}
\label{fig:CrabRelTiming}
\end{figure}

 \ireviewmethree{To validate our local timing analysis, we applied it to observations of the Crab pulsar. This source is frequently observed by \xmm\ to check for any timing inaccuracies} \citep{2012martin}. Its sharp peaks allow a precise measurement of the \gam-\ and X-ray phase lag, which has been observed to remain steady with time \citep{2006kirsch,2010abdocrab}.\\
 We produced three clean event files using the \xmm\ calibration observations performed periodically on the Crab. The first 7.4 ks observation (Obs ID  0611181301, PI Jansen, \ireviewmefour{hereafter observation 1}) was performed on 2012 February 23 (MJD 55981). The second (Obs ID 0611182801, PI Jansen, \ireviewmefour{hereafter observation 2}) was performed on 2015 August 29 (MJD 57263). The third (Obs ID 0611183201, PI Jansen, \ireviewmefour{hereafter observation 3}) was performed on 2016 February 24 (MJD 57442) and lasted 2.9 ks. We selected events from the pn Camera in timing mode, and we barycentered using the {\tt barycen} tool adopting the pulsar position  $\alpha =05^{\rm h}  34^{\rm m} 31\fs97$; $\delta  = +22^\circ 00\arcmin 52\farcs07$ \citep{2008kaplan} and using the DE405 ephemerides.\\
 We then applied the local timing analysis described in Section \ref{sec:relative} to these three observations and \ireviewmefour{fitted the phaseograms with PINT \citep{pint} using two Gaussian peaks in order to evaluate the phase lag between the \gam-ray and X-ray pulse profile}, shown in Fig. \ref{fig:CrabRelTiming}. 
\ireviewmefour{For viewing purposes, we applied a phase offset so that the first peak of the X-ray curve was always located at phase 0.3. Therefore, this example is also useful to illustrate how we define $\phi_{0}$ according to Equation \ref{eq:taylor} and how this does not affect the relative phase difference between X and \gam~phaseograms.}
\ireviewmefour{The phase lag between the \gam\ and X-ray phaseograms for the three \xmm\ observations of the Crab is 0.007 (0.0076$\pm$0.0018 for observation 1, 0.0066$\pm$0.0012 for observation 2, and 0.0063$\pm$0.0018 for observation 3), meaning that the \gam-to-X-ray pulse profile alignment is stable}, in agreement with our expectations.

\end{appendix}

\end{document}